 \documentclass[smallabstract,smallcaptions]{dccpaper}

\usepackage{epsfig}
\usepackage{citesort}
\usepackage{color}
\usepackage{url}

\usepackage{makeidx}  
\usepackage[ruled,vlined,lined,boxed,commentsnumbered]{algorithm2e}
\usepackage{cite}
\usepackage{caption}
\usepackage[usenames,dvipsnames,svgnames,table]{xcolor}
\usepackage[titletoc]{appendix}
\usepackage{graphicx}
\usepackage{wrapfig, blindtext}
\usepackage{amssymb, amsmath} 
\usepackage{mathtools}
\usepackage[margin=1.25in]{geometry}
\usepackage[ruled,vlined,lined,boxed,commentsnumbered]{algorithm2e}
\usepackage{amsthm}

\usepackage{varwidth}
\usepackage{verbatim}

\newtheorem{theorem}{Theorem}
\newtheorem{lemma}{Lemma}

\newenvironment{centerverbatim}{%
	\vspace{10pt}
	\par
	\centering
	\varwidth{\linewidth}%
	\verbatim
}{%
\endverbatim
\endvarwidth
\par
\vspace{10pt}
}

\newcommand{\bigO}{\mathcal{O}}
\newcommand{\dic}{\mathcal D}

\newcommand{\qu}{\mathcal Q}

\newcommand{\extSigma}{\bar\Sigma}

\newlength{\figurewidth}
\newlength{\smallfigurewidth}

\begin{document}

\title
{\large
\textbf{Space-Efficient Re-Pair Compression}
}

\author{%
Philip Bille$^{\ast}$, Inge Li G\o rtz$^{\ast}$, and Nicola Prezza$^{\dag}$\\[0.5em]
{\small\begin{minipage}{\linewidth}\begin{center}
\begin{tabular}{ccc}
$^{\ast}$Technical University of Denmark& \hspace*{0.5in} & $^{\dag}$University of Udine \\
DTU Compute, Asmussens Alle, Bldg. 322  && Via delle scienze, 206 \\
2800 Kgs. Lyngby, Denmark && 33100, Udine, Italy\\
\url{{phbi,inge}@dtu.dk} && \url{prezza.nicola@spes.uniud.it}
\end{tabular}
\end{center}\end{minipage}}
}

\maketitle
\thispagestyle{empty}

\vspace{10pt}
\begin{abstract}
Re-Pair~\cite{larsson2000off} is an effective grammar-based compression scheme achieving strong compression rates in practice. Let $n$, $\sigma$, and $d$ be the text length, alphabet size, and dictionary size of the final grammar, respectively. In their original paper, the authors show how to compute the Re-Pair grammar in expected linear time and $5n + 4\sigma^2 + 4d + \sqrt{n}$ words of working space on top of the text. In this work, we propose two algorithms improving on the space of their original solution. Our model assumes a memory word of $\lceil\log_2 n\rceil$ bits and a re-writable input text composed by $n$ such words. Our first algorithm runs in expected $\bigO(n/\epsilon)$ time and uses 
$(1+\epsilon)n +\sqrt n$ words
of space on top of the text for any parameter $0<\epsilon \leq 1$ chosen in advance. Our second algorithm runs in expected $\bigO(n\log n)$ time and improves the space to 
$n +\sqrt n$ words.
\end{abstract}
\section{Introduction}

Re-Pair (short for recursive pairing) is a grammar-based compression invented in 1999 by Larsson and Moffat~\cite{larsson2000off}. Re-Pair works by replacing a most frequent pair of symbols in the input string by a new symbol, reevaluating the all new frequencies on the resulting string, and then repeating the process until no pairs occur more than once. Specifically, on a string $S$, Re-Pair works as follows. (1) It identifies the most frequent pair of adjacent symbols $ab$. If all pair occur once, the algorithm stops. (2) It adds the rule $A \rightarrow ab$ to the dictionary, where $A$ is a new symbol not appearing in $S$. (3) It repeats the process from step (1). 

Re-Pair achieves strong compression ratios in practice and in theory~\cite{Wan99,GN2007,CN2010, NR2008}. Re-Pair has been used in wide range of applications, e.g., graph representation~\cite{CN2010}, data mining~\cite{TSYYP2016}, and  tree compression~\cite{LMM2013}.

Let $n$, $\sigma$, and $d$ denote the text length, the size of the alphabet, and the size of the dictionary grammar, respectively. Larsson et al.~\cite{larsson2000off} showed how to implement Re-Pair in $O(n)$ expected time and $5n + 4\sigma^2 + 4d + \sqrt{n}$ words of space in addition to the text. \footnote{For simplicity, we ignore any additive $+O(1)$ terms in all space bounds.} The space overhead is due to several data structures used to track the pairs to be replaced and their frequencies. As noted by several authors this makes Re-Pair problematic to apply on large data, and various workarounds have been devised (see e.g.~\cite{Wan99,GN2007,CN2010}). 

Surprisingly, the above bound of the original paper remains the best known complexity for computing the Re-Pair compression. In this work, we propose two algorithms that significantly improve this bound. As in the previous work we assume a standard unit cost RAM with memory words of $\lceil\log_2 n\rceil$ bits and that the input string is given in $n$ such word. Furthermore, we assume that the input string is \emph{re-writeable}, that is, the algorithm is allowed modify the input string during execution, and we only count the space used in addition to this string in our bounds. Since Re-Pair is defined by repeated re-writing operations, we believe this is a natural model for studying this type of compression scheme. Note that we can trivially convert any algorithm with a re-writeable input string to a read-only input string by simply copying the input string to working memory, at the cost of only $n$ extra words of space. We obtain the following result:

\begin{theorem}\label{thm:main}
	Given a re-writeable string $S$ of length $n$ we can compute the Re-Pair compression of $S$ in 
	\begin{itemize}
		\item[](i) $\bigO(n/\epsilon)$ expected time and $(1+\epsilon)n + \sqrt{n}$ words of space for any $0 < \epsilon \leq 1$, or 
		\item[](ii) $\bigO(n\log n)$ expected time and $n + \sqrt{n}$ words.
	\end{itemize}
	
\end{theorem}
Note that since $\epsilon = O(1)$ the time in Thm.~\ref{thm:main}(i) is always at least $\Omega(n)$. For any constant $\epsilon$, (i) matches the optimal linear time bound of Larsson and Moffat~\cite{larsson2000off}, while improving the leading space term by almost $4n$ words to $(1 + \epsilon)n + \sqrt n$ words  (with careful implementation it appears that~\cite{larsson2000off} may be implemented to exploit a re-writeable input string. If so, our improvement is instead almost $3n$ words). Thm.~\ref{thm:main}(ii) further improves the space to $n + \sqrt{n}$ at the cost of increasing time by a logarithmic factor. By choosing $1/\epsilon = o(\log n)$ the time in (i) is faster than (ii) at the cost of a slight increase in space. For instance, with $\epsilon = 1/\log \log n$ we obtain $\bigO(n\log \log n)$ time and $n + n/\log \log n + \sqrt{n}$ words.

 Our algorithm consists of two main phases: high-frequency  and low-frequency pair processing. 
We define a \emph{high-frequency} (resp.\ \emph{low frequency}) pair a character pair appearing at least (resp.\ less than) $\lceil\sqrt{n}/3\rceil$ times in the text (we will clarify later the reason for using constant $3$). Note that there cannot be more than $3\sqrt n$ distinct high-frequency pairs. 
Both phases use two  data structures: a queue $\qu$ storing character pairs (prioritized by frequency) and an array $TP$ storing text positions sorted by character pairs. $\qu$'s elements point to ranges in $TP$ corresponding to all occurrences of a specific character pair.  In Section \ref{sec: sorting pairs} we show how we can sort in-place and in linear time any subset of text positions by character pairs. 
The two phases work exactly in the same way, but  use two different implementations for the queue giving different space/time tradeoffs for operations on it. In both phases, we extract  (high-frequency/low-frequency) pairs from $\qu$ (from the most to least frequent) and replace them in the text with fresh new dictionary symbols. 

When performing a pair replacement $A\rightarrow ab$, for each text occurrence of $ab$ we replace $a$ with $A$ and $b$ with the blank character '\_'. This strategy introduces a potential problem: after several replacements, there could be long (super-constant size) runs of blanks. This could increase the cost of reading pairs in the text by too much. In Section 
\ref{sec:skip blanks} 
we show how we can perform pair replacements while keeping the cost of skipping runs of blanks constant.

\section{Preliminaries}

Let $n$ be the input text's length. Throughout the paper we assume a memory word of size $\lceil\log_2 n\rceil$ bits, and a rewritable input text $T$ on an alphabet $\Sigma$ composed by $n$ such words. In this respect, the working space of our algorithms is defined as the amount of memory used \emph{on top} of the input. Our goal is to minimize this quantity while achieving low running times. For reasons explained later, we reserve two characters (\emph{blank} symbols) denoted as '*' and '\_'. We encode these characters with the integers $n-2$ and $n-1$, respectively \footnote{If the alphabet size is $|\Sigma| < n-1$, then we can reserve the codes $n-2$ and $n-1$ without increasing the number of bits required to write alphabet characters. Otherwise, if $|\Sigma| \geq n-1$ note that the two (or one) alphabet characters with codes $n-2 \leq x,y \leq n-1$ appear in at most two text positions $i_1$ and $i_2$, let's say $T[i_1]=x$ and $T[i_2]=y$. Then, we can overwrite $T[i_1]$ and $T[i_2]$ with the value 0 and store separately two pairs $\langle i_1,x \rangle$, $\langle i_2,y \rangle$. Every time we read a value $T[j]$ equal to $0$, in constant time we can discover whether  $T[j]$ contains $0$, $x$, or $y$. 
Throughout the paper we will therefore assume that $|\Sigma|\leq n$ and that characters from $\Sigma\cup\{*,\_\}$ fit in $\lceil\log_2 n\rceil$ bits.}.

The Re-Pair compression scheme works by replacing character pairs (with frequency at least 2) with fresh new symbols. We use the notation $\dic$ to indicate the \emph{dictionary} of such new symbols,  and denote by $\extSigma$ the extended alphabet $\extSigma = \Sigma \cup \dic$.
It is easy to prove (by induction) that $|\extSigma| \leq n$: it follows that we can fit both alphabet characters and dictionary symbols in $\lceil\log_2 n\rceil$ bits.
The output of our algorithms consists in a set of rules of the form $X \rightarrow AB$, with $A,B\in\extSigma$ and $X\in\dic$. Our algorithms stream the set of rules directly to the output (e.g.\ disk), so we do not count the space to store them in main memory.

\section{Main Algorithm}

We describe our strategy top-down: first, we introduce the queue $\qu$ as a blackbox, and use it to describe our main algorithm. In the next sections we describe the high-frequency and low-frequency pair processing queues implementations.

\subsection{The queue as a blackbox}

Our queues support the following operations:\\\ \\
- $\mathbf{new\_low\_freq\_queue(T,TP)}$. Return the  low-frequency pairs queue.\\
- $\mathbf{new\_high\_freq\_queue(T,TP)}$. Return the  high-frequency pairs queue.\\
- $\mathbf{\qu[ab]},\ ab\in \extSigma^2$. If $ab$ is in the queue, return a triple $\langle P_{ab}, L_{ab}, F_{ab} \rangle$, with $L_{ab}\geq F_{ab}$ such that:

(i) $ab$ has frequency $F_{ab}$ in the text.

(ii) All text occurrences of $ab$ are contained in $TP[P_{ab},\dots , P_{ab}+L_{ab}-1]$.\\
- $\mathbf{\qu.max()/\qu.min()}$: return the pair $ab$  in $\qu$ with the highest/lowest $F_{ab}$.\\
- $\mathbf{\qu.remove(ab)}$: delete $ab$ from $\qu$.\\
- $\mathbf{\qu.contains(ab)}$: return true iff $\qu$ contains pair $ab$.\\
- $\mathbf{\qu.size()}$ return the number of pairs stored in $\qu$.\\
- $\mathbf{\qu.decrease(ab)}$: decrease $F_{ab}$ by one.\\
- $\mathbf{\qu.synchronize(AB)}$. If $F_{AB}<L_{AB}$, then  $TP[P_{AB},\dots , P_{AB}+L_{AB}-1]$ contains occurrences of pairs $XY\neq AB$ (and/or blank positions). The procedure sorts $TP[P_{AB},\dots , P_{AB}+L_{AB}-1]$ by character pairs (ignoring positions containing a blank) and, for each such $XY$, removes the least frequent pair in $\qu$ and creates a new queue element for $XY$ pointing to the range in $TP$ corresponding to the occurrences of $XY$. If  $XY$ is less frequent than the least frequent pair in $\qu$, $XY$ is not inserted in the queue. Before exiting, the procedure re-computes $P_{AB}$ and $L_{AB}$ so that $TP[P_{AB},\dots , P_{AB}+L_{AB}-1]$ contains all and only the occurrences of $AB$ in the text (in particular, $L_{AB}=F_{AB}$).


\subsection{Algorithm description}

In Algorithm \ref{alg:substitution round} we describe the procedure substituting the most frequent pair in the text with a fresh new dictionary symbol. We use this procedure in Algorithm \ref{alg:repair} to compute the re-pair grammar. Variables $T$ (the text), $TP$ (array of text positions), and $X$ (next free dictionary symbol) are global, so we do not pass them from Algorithm \ref{alg:repair} to Algorithm \ref{alg:substitution round}. Note that---in Algorithm \ref{alg:substitution round}---new pairs appearing after a substitution can be inserted in $\qu$ only inside procedure $\qu.synchronize$ at Lines  \ref{line:synchro1}, and \ref{line:synchro2}. However, operation at Line \ref{line:synchro1} is executed only under a certain condition. As discussed in the next sections, this trick allows us to amortize operations while preserving correctness of the algorithm. 

In Lines \ref{line:for1}, \ref{line:context1}, and \ref{line:context2} of Algorithm \ref{alg:substitution round} we assume that---if necessary---we are skipping runs of blanks while extracting text characters (constant time, see Section \ref{sec:skip blanks}). In Line \ref{line:context1} we extract $AB$ and the two symbols $x,y$ preceding and following it (skipping runs of blanks if necessary). In Line \ref{line:context2}, we extract a text substring $s$ composed by $X$ and the symbol preceding it (skipping runs of blanks if necessary). After this, we replace each $X$ with $AB$ in $s$ and truncate $s$ to its suffix of length 3. This is required since we need to reconstruct $AB$'s context \emph{before} the replacement took place. Moreover, note that the procedure could return $BAB$ if we replaced a substring $ABAB$ with $XX$.

\vspace{10pt}

\begin{algorithm}[th!]
	\caption{$substitution\_round(\qu)$}
	\label{alg:substitution round}
	
	\SetKwInOut{Input}{input}
	\SetKwInOut{Output}{behavior}
	\SetSideCommentLeft
	\LinesNumbered
	
	\Input{The queue $\qu$}
	\Output{Pick the most frequent pair from $\qu$ and replace its occurrences in the text with a new dictionary symbol}
	
	\BlankLine
	
	$AB \leftarrow \qu.max()$\;\label{line:pick max}
	$\overline{ab} \leftarrow \qu.min()$\tcc*[r]{global variable storing least frequent pair}\label{line:less freq pair}
	
	\BlankLine
	
	$\mathbf{output}\ X\rightarrow AB$\tcc*[r]{output new rule}
	
	\BlankLine				
	
	\For{$i=TP[P_{AB}],\dots, TP[P_{AB}+L_{AB}-1]$ $\mathbf{and}$ $T[i,i+1]=AB$}{\label{line:for1}
		
		\BlankLine	
		
		$xABy\leftarrow get\_context(T,i)$\tcc*[r]{$AB$'s context (before replacement)}\label{line:context1}
		
		$replace(T,i,X)$\tcc*[r]{Replace $X\rightarrow AB$ at position $i$ in $T$}
		
		\BlankLine	
		
		\If{$\qu.contains(xA)$}{
			
			$\qu.decrease(xA)$\;\label{line:decrease1}
			
		}
		
		\BlankLine
		
		\If{$\qu.contains(By)$}{
			
			$\qu.decrease(By)$\;\label{line:decrease2}
			
		}

	}
	
	\For{$i=TP[P_{AB}],\dots, TP[P_{AB}+L_{AB}-1]$ $\mathbf{and}$ $T[i]=X$}{
		
		\BlankLine	
		
		$xAB\leftarrow get\_context'(T,i)$\tcc*[r]{$X$'s left context}\label{line:context2}
		
		\BlankLine	
		
		\If{$\qu.contains(xA)$ $\mathbf{and}$ $F_{xA}\leq L_{xA}/2$}{\label{line:amortize}
			
			$\qu.synchronize(xA)$\;\label{line:synchro1}

		}
		
		%
		%
		%
		%
		
	}
	
	\BlankLine
	$\qu.synchronize(AB)$\tcc*[r]{Find new pairs in $AB$'s occurrences list}\label{line:synchro2}
	
	$\mathcal \qu.remove(AB)$\;
	$X \leftarrow X+1$\tcc*[r]{New dictionary symbol}
	\BlankLine
	
\end{algorithm}

\begin{algorithm}[th!]
	\caption{$compute\_repair(T)$}
	\label{alg:repair}
	
	\SetKwInOut{Input}{input}
	\SetKwInOut{Output}{behavior}
	\SetSideCommentLeft
	\LinesNumbered
	
	\Input{Text $T\in\Sigma^n$}
	\Output{The re-pair grammar of $T$ is computed and streamed to output}
	
	\BlankLine
	
	\BlankLine
	
	$n\leftarrow |T|$\;
	$X\leftarrow |\Sigma|$\tcc*[r]{next dictionary symbol}
	
	\While{$highest\_frequency(T)\geq \sqrt n/3$}{\label{line:while1}
		
		\BlankLine
		
		$TP \leftarrow sort\_pairs(T)$\tcc*[r]{sort $T$'s positions by pairs}
		
		$\qu \leftarrow new\_high\_freq\_queue(T,TP)$\;
		
		\BlankLine
		
		\While{$\qu.size()>0$}{
			
			$substitution\_round(\qu)$\;
			
		}	
		
		\BlankLine
		
		$free\_memory(\qu, TP)$\;
		$compact\_text(T)$\tcc*[r]{delete blanks}\label{line:compact text1}
		
	}

	\While{$highest\_frequency(T) > 2$}{\label{line:while2}
		
		\BlankLine
		
		$TP \leftarrow sort\_pairs(T)$\tcc*[r]{sort $T$'s positions by pairs}\label{line:sort pairs2}
		
		$\qu \leftarrow new\_low\_freq\_queue(T,TP)$\;\label{line:new LF queue}
		
		\BlankLine
		
		\While{$\qu.size()>0$}{
			
			$substitution\_round(\qu)$\;
			
		}	
		
		\BlankLine
		
		$free\_memory(\qu, TP)$\;\label{line:free mem2}
		$compact\_text(T)$\tcc*[r]{delete blanks}\label{line:compact text2}
		
	}

\end{algorithm}

\subsection{Amortization: correctness and complexity}\label{sec:amortization}

Assuming the correctness of the queue implementations (see next sections), all we are left to show is the correctness of our amortization policy at Lines \ref{line:amortize} and \ref{line:synchro1} of Algorithm \ref{alg:substitution round}. More formally: in Algorithm \ref{alg:substitution round}, replacements create new pairs; however, to amortize operations we postpone the insertion of such pairs in the queue (Line \ref{line:synchro1} of Algorithm \ref{alg:substitution round}). To prove the correctness of our algorithm, we need to show that every time we pick the maximum $AB$ from $\qu$ (Line \ref{line:pick max}, Algorithm \ref{alg:substitution round}), $AB$ is the pair with the highest frequency in the text (i.e.\ all postponed pairs have lower frequency than $AB$). 
Suppose, by contradiction, that at Line \ref{line:pick max} of Algorithm \ref{alg:substitution round} we pick pair $AB$, but the highest-frequency pair in the text is $CD\neq AB$. Since $CD$ is not in $\qu$, we have that (i) $CD$ appeared after some substitution $D\rightarrow zw$ which generated occurrences of $CD$ in portions of the text containing $Czw$, and\footnote{Note that, if $CD$ appears after some substitution $C\rightarrow zw$ which creates occurrences of $CD$ in portions of the text containing $zwD$, then all occurrences of $CD$ are contained in $TP[P_{zw},\dots,P_{zw}+L_{zw}-1]$, and we insert $CD$ in $\mathcal Q$ at Line \ref{line:synchro2} of Algorithm \ref{alg:substitution round} within procedure $\qu.synchronize(zw)$} (ii) $F_{Cz}>L_{Cz}/2$, otherwise the synchronization step at Line \ref{line:synchro1} of Algorithm \ref{alg:substitution round} ($\qu.synchronize(Cz)$) would have been executed,  and $CD$ would have been inserted in $\qu$. Note that all occurrences of $CD$ are contained in $TP[P_{Cz},\dots,P_{Cz}+L_{Cz}-1]$.
$F_{Cz} > L_{Cz}/2$ means that \emph{more than half} of the entries $TP[P_{Cz},...,P_{Cz}+L_{Cz}-1]$ contain an occurrence of $Cz$, which implies than \emph{less than half} of such entries contain occurrences of pairs different than $Cz$ (in particular $CD$, since $D\neq z$). This, combined with the fact that all occurrences of $CD$ are stored in $TP[P_{Cz},...,P_{Cz}+L_{Cz}-1]$, yields $F_{CD} \leq L_{Cz}/2$. Then, $F_{CD} \leq L_{Cz}/2 < F_{Cz}$ means that $Cz$ has a higher frequency than $CD$. This leads to a contradiction, since we assumed that $CD$ was the pair with the highest frequency in the text.

Note that operation $\qu.synchronize(xA)$ at Line \ref{line:synchro1} of Algorithm \ref{alg:substitution round} scans $xA$'s occurrences list ($\Theta(L_{xA})$ time). However, to keep time under control, in Algorithm \ref{alg:substitution round} we are allowed to spend only time proportional to $F_{AB}$. Since $L_{xA}$ could be much bigger than $F_{AB}$, we need to show that our strategy amortizes operations. Consider an occurrence $xABy$ of $AB$ in the text. After replacement $X\rightarrow AB$, this text substring becomes $xXy$. In Lines \ref{line:decrease1}-\ref{line:decrease2} we decrease by one in constant time the two frequencies $F_{xA}$ and $F_{By}$ (if they are stored in $\qu$). Note: we manipulate just $F_{xA}$ and $F_{By}$, and not the actual intervals associated with these two pairs. As a consequence, for a general pair $ab$ in $\qu$, values $F_{ab}$ and $L_{ab}$ do not always coincide. However, we make sure that, when calling $\qu.max()$ at Line \ref{line:pick max} of Algorithm \ref{alg:substitution round}, the following invariant holds for every pair $ab$ in the priority queue:
$$
F_{ab} > L_{ab}/2
$$
The invariant is maintained by calling $\qu.synchronize(xA)$ (Line \ref{line:synchro1}, Algorithm \ref{alg:substitution round}) as soon as we decrease by ``too much'' $F_{xA}$ (i.e.\ $F_{xA} \leq L_{xA}/2$). 
It is easy to see that this policy amortizes operations: every time we call procedure $\qu.synchronize(ab)$, either---Line \ref{line:synchro2}---we are replacing $ab$ with a fresh new dictionary symbol (thus $L_{ab} < 2\cdot F_{ab}$ work is allowed), or---Line \ref{line:synchro1}---we just decreased $F_{ab}$ by too much ($F_{ab} \leq L_{ab}/2$). In the latter case, we already have done at least $L_{ab}/2$ work during previous replacements (each one has decreased $ab$'s frequency by 1), so $\bigO(L_{ab})$ additional work does not asymptotically increase running times.

\section{Details and Analysis}

We first describe how we implement character replacement in the text and how we efficiently sort text positions by pairs. Then, we provide the two queue implementations. For the low-frequency pairs queue, we provide two alternative implementations leading to two different space/time tradeoffs for our main algorithm. We conclude by analyzing the complexity of Algorithm \ref{alg:repair} with the different queue implementations.

\subsection{Skipping blanks in constant time}\label{sec:skip blanks}

As noted above, pair replacements generate runs of the blank character '\_'. Our aim in this section is to show how to skip these runs in constant time. Recall that the text is composed by $\lceil\log_2n\rceil$-bits words. 
%
%
Recall that we reserve \emph{two} blank characters: '*' and '\_'. If the run length $r$ satisfies $r<10$, then we fill all run positions with character '\_' (skipping this run takes constant time). Otherwise, ($r\geq 10$) we start and end the run with the string \texttt{\_*i*\_}, where $i=r-1$, and fill the remaining run positions with '\_'. For example, the text \texttt{aB\_\_\_\_\_\_\_\_\_\_\_c} is stored as
\begin{centerverbatim}
	a B _ * 10 * _ _ _ * 10 * _ c
\end{centerverbatim}
Note that, with this solution, only run lengths are delimited by character '*': it follows that we can distinguish between run lengths and alphabet characters. We remind the reader that we encode the extra characters '*' and '\_' with the integers $n-2$ and $n-1$, respectively. Then, it is easy to see that any integer $i$ storing a run length (minus 1) always satisfies $i\leq n-3$, so we can safely distinguish between run lengths and reserved blank characters.
Our text representation is completely transparent in that it allows to retrieve any text character/blank in constant time: if $T[j]>n-3$, then $T[j]$ contains a blank. If $T[j]\leq n-3$, if $T[j-1]=T[j+1]=n-2$ then $T[j]$ contains a blank, otherwise an alphabet character.

The only thing we are left to show is how to merge two runs of blanks in the case a single alphabet character between them is replaced by a blank after a substitution. For example, the text
\begin{centerverbatim}
	a B _ * 10 * _ _ _ * 10 * _ C _ * 11 * _ _ _ _ * 11 * _ D
\end{centerverbatim}
after substitution $E \rightarrow BC$ should become
\begin{centerverbatim}
	a E _ * 23 * _ _ _ _ _ _ _ _ _ _ _ _ _ _ _ _ * 23 * _ D
\end{centerverbatim}
It is easy to see that the replacement can be implemented in constant time starting from the position containing 'B', so we do not discuss it further. 
In the paper we assume that the text is stored with the above representation, with constant-time cost for skipping runs of blanks.

\subsection{Sorting pairs and frequency counting}\label{sec: sorting pairs}
Let $T$ be an array of $n$ entries each consisting of a word of $\log n$ bits (for simplicity, we assume that $n$ is a power of two). In this section we show how to sort the pairs in $T$ lexicographically in linear time using additional 
$n$ words
of memory. Our algorithm only requires read-only access to $T$. Furthermore, the algorithm generalizes substrings of any constant length in the same complexity. As an immediate corollary, we can compute the frequency of each pair in the same complexity simply by traversing the sorted sequence.  

Our solution needs the following results on in-place sorting and merging. 

\begin{lemma}[Franceschini et al. \cite{FMP2007}]\label{lem:inplacesort}
	Given an array $A$ of length $n$ with $\bigO(\log n)$ bit entries, we can in-place sort $A$ in $\bigO(n)$ time. 
\end{lemma}

\begin{lemma}[Salowe and Steiger\cite{SS1987}]\label{lem:inplacemerge}
	Given arrays $A$ and $B$ of total length $n$, we can merge $A$ and $B$ in-place using a comparison-based algorithm in $\bigO(n)$ time. 
\end{lemma}

The above result immediately provides simple but inefficient solutions to sorting pairs. In particular, we can copy each pair of $T$ into an array of $n$ entries each storing a pair using $2$ words, and then in-place sort the array using Lemma~\ref{lem:inplacesort}. This uses  $\bigO(n)$ time but requires $2 n$ words space. Alternatively, we can copy the positions of each pair into an array and then apply a comparison-based in-place sorting algorithm based on~\ref{lem:inplacemerge}. This uses $\bigO(n \log n)$ time but only requires 
$n$ words
of space. Our result simultaneously  obtains the best of these time and space bounds.

Our algorithm works as follows. Let $A$ be an array of $n$ words. We greedily process $T$ from left-to-right in phases. In each phase we process a contiguous segment $T[i, j]$ of overlapping pairs of $T$ and compute and store the corresponding sorted segment in $A[i,j]$. Phase $i = 0, \ldots, k$ proceeds as follows. Let $r_i$ denote the number of \emph{remaining pairs} in $T$ not yet processed. Initially, we have that $r_0 = n$. Note that $r_i$ is also the number of unused entries in $A$. We copy the next $r_i/3$ pairs of $T$ into $A$. Each pair is encoded using the two characters of the pair and the position of the pair in $T$. Hence, each encoded pair uses $3$ words and thus fills all remaining $r_i$ entries in $A$. We sort the encoded segment using the in-place sort from Lemma~\ref{lem:inplacesort}, where each $3$-words encoded pair is viewed as a single key. We then compact the segment back into $r_i/3$ only entries of $A$ by throwing away the characters of each pair and only keeping the position of the pair. We repeat the process until all pairs in $T$ have been processed. At the end $A$ consists of a collection of segments of sorted pairs. We merge the segments from right-to-left using the in-place comparison-based merge from Lemma~\ref{lem:inplacemerge} (note that segment borders can be detected by accessing the text, so we do not need to store them separately).

Next we analyse the algorithm. For the space bound, note that at any point in time the algorithm never uses more than $\bigO(1)$ words in addition to the remaining entries in $A$. Hence, the total space is 
$n+\bigO(1)$ words.
For the time bound, note that each phase decreases the number of remaining pairs in $A$ by a third of the current number of remaining pairs. Hence, for $i > 0$, $r_{i} = r_{i-1} - r_{i-1}/3 = (2/3)r_i$. Since $r_0 = n$ it follows that $r_{i} = (2/3)^i n$. The total number of phases is thus $k=\log_{3/2} n$. By Lemma~\ref{lem:inplacesort} phase $i$ uses $\bigO(r_i)$ time and hence the total time for all phases is $\sum_{i}^k \bigO(r_i) = \bigO(n)$. For the merging step, note that the last step merges two segments of total size $n$, the second last step merges two segments of total size $(2/3) n$, and in general the $i$th last step merges two segments of total size $(2/3)^i n$. By Lemma~\ref{lem:inplacemerge} the total time is thus $\sum_{i}^k \bigO((2/3)^i n) = \bigO(n)$. In summary, we have the following result for sorting and immediately corollary for counting frequencies.

\begin{lemma}
	Given a string $T$ of length $n$ with $\lceil\log_2 n\rceil$-bit characters, we can sort the pairs of $T$ in $\bigO(n)$ time using 
	$n+\bigO(1)$ words.
\end{lemma}

\begin{lemma}\label{lemma:count}
	Given a string $T$ of length $n$ with $\lceil\log_2 n\rceil$-bit characters, we can count the frequencies of pairs of $T$ in $\bigO(n)$ time using 
	$n+\bigO(1)$ words.
\end{lemma}

We use Lemma \ref{lemma:count} to find the maximum frequency in linear time and $n$ words of space in Algorithm \ref{alg:repair} (procedure $highest\_frequency(T)$).
Note that our technique can be used to sort in-place and linear time any subset of text positions by character pairs (required in our queue implementations). Finally, note that with our strategy text positions are sorted first by character pairs, and then by position (this follows from the fact that---inside our sorting procedure---we concatenate the pair and the text position in a single integer of 3 words). This fact is important in our algorithm since it allows us to process text pairs from left to right.

\subsection{High-Frequency Pairs Queue}\label{sec:hf pairs}

The capacity of the high-frequency pairs queue is $\sqrt n/11$. We implement $\qu$ with the following two components:\\\ \\
(i) \textbf{Hash} $\mathcal H$. We keep a hash table $\mathcal H:\extSigma^2 \rightarrow [0,\sqrt n/11]$ with  $(2/11)\sqrt n$ entries. $\mathcal H$ will be filled with at most  $\sqrt n/11$ pairs (hash load $\leq 0.5$). Collisions are solved by linear probing. The overall size of the hash is $(6/11)\sqrt n$ words: 3 words (one pair and one integer) per hash entry.\\	
(ii) \textbf{Queue array} $B$. We keep an  array $B$ of quadruples from  $\extSigma^2 \times [0,n) \times [0,n) \times [0,n)$. $B$ will be filled with at most $\sqrt n/11$ entries. 
We denote with $\langle ab, P_{ab}, L_{ab}, F_{ab} \rangle$ a generic element of $B$. The idea is that $B$ stores the most frequent character pairs, together with their frequencies.
Every time we pop the highest-frequency pair from the queue, the following holds: (i) $ab$ has frequency $F_{ab}$ in the text, and (ii) $ab$ occurs in a \emph{subset} of text positions $TP[P_{ab}], \dots, TP[P_{ab}+L_{ab}-1]$. The overall size of $B$ is $(5/11)\sqrt n$ words.\\

\begin{figure}[h!]
	\begin{center}
		\includegraphics[trim=2cm 15cm 1cm 1cm, clip=true, width=0.7\textwidth]{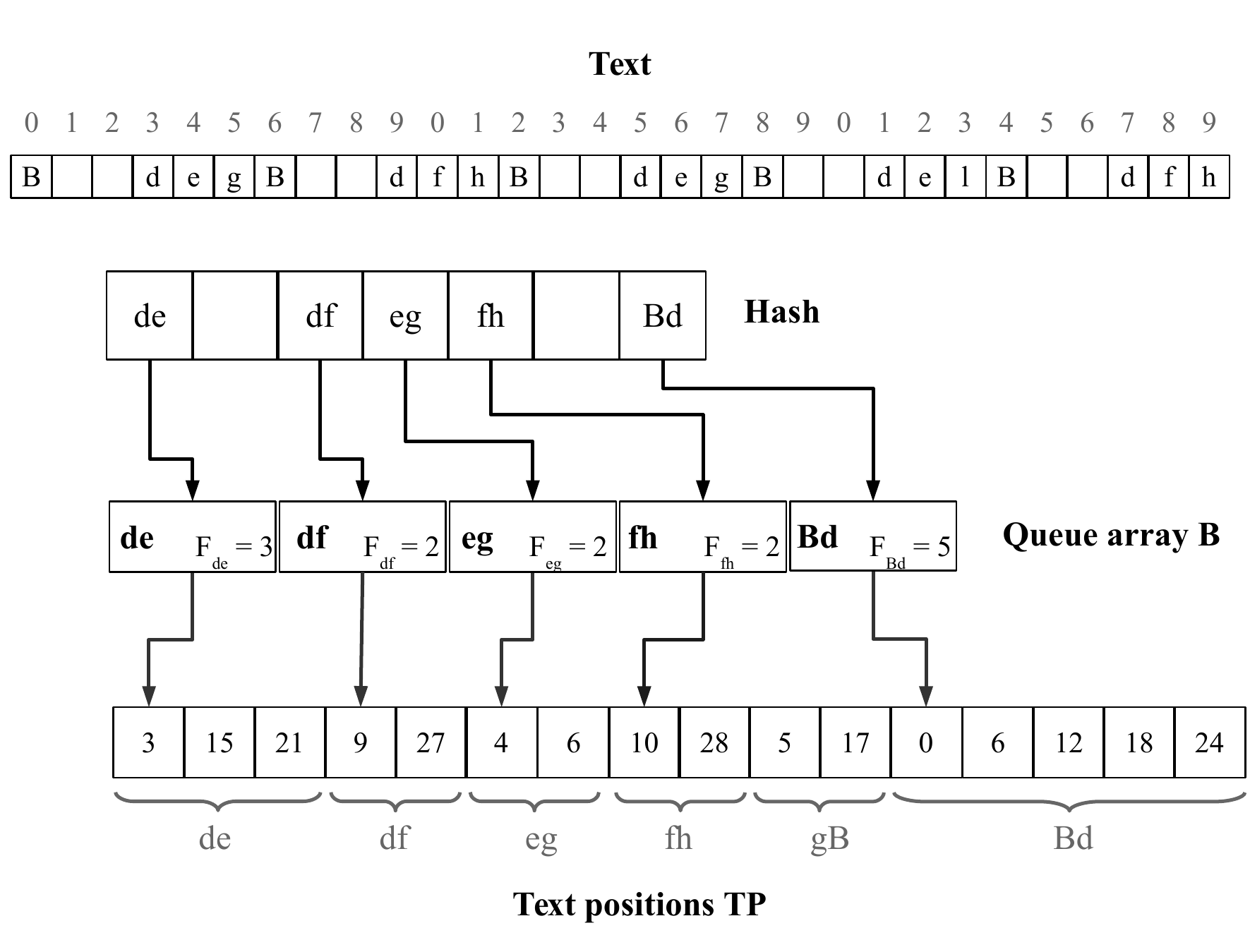}
	\end{center}
	\caption{
		The high-frequency queue (hash and queue array). In this example, the text is obtained from \texttt{abcdegabcdfhabcdegabcdelabcdfh} after the substitutions (in order) $A\rightarrow ab$ and $B\rightarrow Ac$. Pairs $de$, $df$, $eg$, $fh$, and $Bd$ have been inserted in the queue. In this example, the $L$-field of $B$ elements is always equal to the $F$-field, so we do not show it to improve readability.
		\vspace{10pt} 
	}\label{fig: HF}
\end{figure}

$\mathcal H$'s entries point to $B$'s entries: at any stage of the algorithm, if $\mathcal H$ contains a pair $ab$, then $B[\mathcal H[ab]] = \langle ab, P_{ab}, L_{ab}, F_{ab} \rangle$ is the quadruple associated with the pair. 
Overall, $\qu=\langle\mathcal H,B\rangle$ takes $\sqrt n$ words of space. Figure \ref{fig: HF} depicts our high-frequency queue.

First, we show how function $new\_high\_freq\_queue(T,TP)$ is implemented. We scan $TP$ from left to right and replace every maximal sub-array $TP[i,\dots, i+k-1]$, $k\geq 2$, corresponding to a character pair $T[TP[i],TP[i]+1]$ with the integers pair $\langle k,TP[i] \rangle$. We store such pair in two words by concatenating the integers $k$ and $TP[i]$. Whenever $k=1$ or $k>2$, we compact $TP$ positions so that all pairs are stored consecutively. In the end, (an opportune prefix of) $TP$ contains a list $\langle k_1,j_1 \rangle, \dots, \langle k_t,j_t \rangle$ of pairs representing the frequency of all character pairs with frequency at least $2$: $\langle k,j \rangle$ is in this list iff pair $T[j] T[j+1]$ appears $k\geq 2$ times in the text. We conclude by sorting this list in decreasing frequency order with the Algorithm described in~\cite{FMP2007}. Then, we scan this list left-to-right and insert at most $\sqrt n/11$ high-frequent pairs in $\mathcal H$, starting from the most frequent ones. Finally, we re-build $TP$ (re-using the memory already allocated for it), and scan it left-to-right. For each maximal $TP[p,\dots,p+k-1]$ corresponding to a pair $ab$, if $ab$ is in $\mathcal H$ then we set $\mathcal H[ab]\leftarrow |B|$ and append $\langle ab,p,k,k \rangle$ at the end of $B$. This procedure runs in $\bigO(n)$ time and uses $\bigO(1)$ words of space in addition to $T$, $TP$, $B$, and $\mathcal H$.
Operations on $\qu$ are implemented as follows:

\begin{itemize}
	\item $\qu[ab],\ ab\in \extSigma^2$: return the last three components of the quadruple $B[\mathcal H[ab]]$. $\bigO(1)$ expected time.
	\item $\qu.max()$ is implemented by scanning $B$. $\bigO(\sqrt n)$ time. 
	\item $\qu.min()$ is implemented by scanning $B$. $\bigO(\sqrt n)$ time. 
	\item $\qu.remove(ab)$: set $B[\mathcal H[ab]]\leftarrow \langle NULL,NULL,NULL,NULL\rangle$  and delete $ab$ from $\mathcal H$. $\bigO(1)$ expected time.
	\item $\qu.contains(ab)$. This operation requires one access to $\mathcal H$. $\bigO(1)$ expected time.
	\item $\qu.size()$: we only need to keep a variable storing the current size and update it at each remove operation. $\bigO(1)$ time.
	\item $\qu.decrease(AB)$: decrease the fourth component of $B[\mathcal H[AB]]$ by one. If $F_{AB}$ becomes smaller than $\lceil\sqrt{n}/3\rceil$, remove the pair from the queue. $\bigO(1)$ expected time.
	\item $\qu.synchronize(AB)$. Let $\overline{ab}$ be a global variable storing the pair with the smallest frequency in $\qu$ (Algorithm \ref{alg:substitution round}, Line \ref{line:less freq pair}).
	We apply the sorting algorithm of Section \ref{sec: sorting pairs} to sub-array $TP'=TP[P_{AB},\dots,P_{AB}+L_{AB}-1]$, excluding text positions that contain a blank character. After this, positions are clustered in $TP'$ by character pairs. For each contiguous maximal sub-array  $TP[p,\dots, p+k-1]$ of $TP'$ corresponding to a pair $XY$:
	
	\begin{itemize}
		\item If $XY\neq AB$ and $k > F_{\overline{ab}}$, we remove $\overline{ab}$ from $\qu$ and insert $XY$ as follows. Let $j_{min} = \mathcal{H}[\overline{ab}]$. We remove $\overline{ab}$ from the hash $\mathcal H$, overwrite  $B[j_{min}] \leftarrow \langle XY, p, k, k \rangle$, and insert $\mathcal H[XY] \leftarrow j_{min}$ in the hash. Then, we recompute $\overline{ab}=\qu.min()$. 
		
		\item If $XY= AB$, then we just overwrite  $B[\mathcal H[AB]] \leftarrow \langle AB, p, k, k \rangle$ and set $\overline{ab}\leftarrow AB$ if $k<F_{\overline{ab}}$.
	\end{itemize}
	
	Running time: $\bigO(L_{AB} + N\cdot \sqrt n)$, where $L_{AB}$ is $AB$'s interval length at the moment of entering in this procedure, and $N$ is the number of new pairs $XY$ inserted in the queue (we call $\qu.min()$ for each of them).
	
\end{itemize} 
\textbf{Time complexity} Note that to find the most frequent pair in $\qu$ we scan all $\qu$'s elements; since $|\qu|\in\bigO(\sqrt n)$ and there are at most $3\sqrt n$ high-frequency pairs, the overall time spent inside procedure $max(\qu)$ does not exceed $\bigO(n)$. 

Since we insert at most $\sqrt n/11$ pairs in $\qu$ but there may be up to $3\sqrt n$ high-frequency pairs, once $\qu$ is empty we may need to fill it again with new high-frequency pairs (\texttt{while} loop at line \ref{line:while1} of Algorithm \ref{alg:repair}). We need to repeat this process at most $(3\sqrt n)/(\sqrt n/11)\in\bigO(1)$ times, so the number of rounds is constant.

We call $\qu.min()$ in two cases: (i) after extracting the maximum from $\qu$ (Line \ref{line:less freq pair}, Algorithm \ref{alg:substitution round}), and (ii) within procedure $\qu.synchronize$, after discovering a new high-frequency pair $XY$ and inserting it in $\qu$. Case (i) cannot happen more than $3\sqrt n$ times. As for case (ii), note that a high-frequency pair can be inserted at most once per round in $\qu$ within procedure $\qu.synchronize$. Since the overall number of rounds is constant and there are at most $3\sqrt n$ high-frequency pairs, also in this case we call $\qu.min()$ at most $\bigO(\sqrt n)$ times. Overall, the time spent inside $\qu.min()$ is therefore $\bigO(n)$.

Finally, considerations of Section \ref{sec:amortization} imply that scanning/sorting occurrences lists inside operation $\qu.synchronize$ take overall linear time thanks to our amortization policy.

\subsection{Low-Frequency Pairs Queue}

We describe two low-frequency queue variants, denoted in what follows as \emph{fast} and \emph{light}. We start with the fast variant.\\\ \\
\textbf{Fast queue} Let $0<\epsilon \leq 1$ be a parameter chosen in advance. Our fast queue has maximum capacity $(\epsilon/13)\cdot n$ and is implemented with three components:\\\ \\
(i) \textbf{Set of doubly-linked lists} $B$. This is a set of lists; each list is associated to a distinct frequency. $B$ is implemented as an array of elements of the form $\langle ab, P_{ab}, L_{ab}, F_{ab}, Prev_{ab}, Next_{ab} \rangle$, where:\\
- $ab, P_{ab}, L_{ab}$, and $F_{ab}$ have the same meaning as in the high-frequency queue\\
- $Prev_{ab}$ points to the previous $B$ element with frequency $F_{ab}$ (NULL if this is the first such element)\\
- $Next_{ab}$ points to the next $B$ element with frequency $F_{ab}$ (NULL if this is the last such element)\\
Every $B$ element takes 7 words. We allocate $\epsilon\cdot(7/13)\cdot n$ words for $B$ (maximum capacity: $(\epsilon/13)\cdot n$)\\ 
(ii) \textbf{Doubly-linked frequency vector} $\mathcal F$. This is a word vector $\mathcal F[0,\dots, \sqrt n/3-1]$ indexing all possible frequencies of low-frequency pairs. We say that $\mathcal F[i]$ is \emph{empty} if $i$ is not the frequency of any pair in $T$. In this case, $\mathcal F[i]=NULL$. Non-empty $\mathcal F$'s entries are  doubly-linked: we associate to each $\mathcal F[i]$ two values $\mathcal F[i].prev$ and $\mathcal F[i].next$ representing the two non-empty pair's frequencies immediately smaller/larger than $i$. We moreover keep two variables $MAX$ and $MIN$ storing the largest and smallest frequencies in $\mathcal F$. If $i$ is the frequency of some character pair, then $\mathcal F[i]$ points to the first $B$ element in the chain associated with frequency $i$: $B[\mathcal F[i]] = \langle ab, P_{ab}, L_{ab}, i, NULL, Next_{ab} \rangle$, for some pair $ab$. $\mathcal F$ takes overall $\sqrt n$ words of space\\	
(iii) \textbf{Hash} $\mathcal H$. We keep a hash table $\mathcal H:\Sigma^2 \rightarrow [0,n]$ with  $\epsilon\cdot (2/13)\cdot n$ entries. The hash is indexed by character pairs. $\mathcal H$ will be filled with at most  $\epsilon\cdot n/13$ pairs (hash load $\leq 0.5$). Collisions are solved by linear probing. The overall size of the hash is $\epsilon\cdot (6/13)\cdot n$ words: 3 words (one pair and one integer) per hash entry.
$\mathcal H$'s entries point to $B$'s entries: if $ab$ is in the hash, then $B[\mathcal H[ab]]= \langle ab, P_{ab}, L_{ab}, F_{ab}, Prev_{ab}, Next_{ab} \rangle$\\

\begin{figure}[h!]
	\begin{center}
		\includegraphics[trim=0cm 9cm 0cm 1cm, clip=true, width=0.7\textwidth]{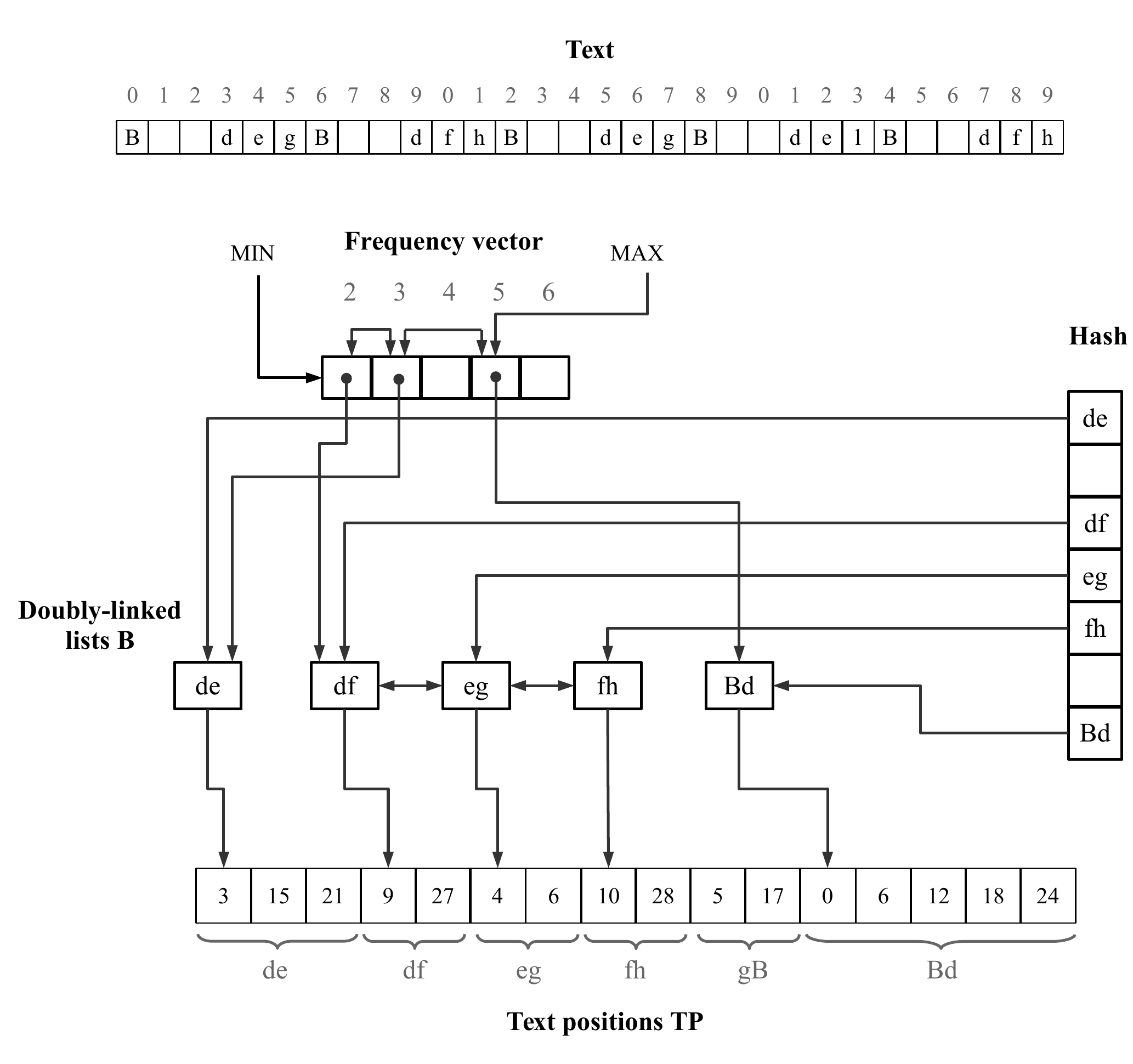}
	\end{center}
	\caption{
		The low-frequency queue is composed by three components: frequency vector, hash, and doubly-linked lists. In this example, the text is obtained from \texttt{abcdegabcdfhabcdegabcdelabcdfh} after the substitutions (in order) $A\rightarrow ab$ and $B\rightarrow Ac$. Pairs $de$, $df$, $eg$, $fh$, and $Bd$ have been inserted in the queue. We do not show $F$- and $L$-fields in $B$'s elements to improve readability.\vspace{10pt}
	}\label{fig: LF}
\end{figure}

Overall, $\qu = \langle B,\mathcal F, \mathcal H \rangle$ takes $\sqrt n + \epsilon\cdot n$ words of space. Figure \ref{fig: LF} depicts our low-frequency queue.

First, we show how function $new\_low\_freq\_queue(T,TP)$ is implemented. We build list $\langle k_1,j_1 \rangle, \dots, \langle k_t,j_t \rangle$ sorted by frequency as done for the high-frequency queue. We scan this list left-to-right and insert at most $n/13$ low-frequency pairs in $\mathcal H$, starting from the most frequent ones. At the same time, we fill $\mathcal F$'s list pointers: while processing $\langle k_i,j_i \rangle$, if the maximum queue capacity has not been reached then we set $\mathcal F[k_{i-1}].next \leftarrow k_i$, and $\mathcal F[k_{i}].prev \leftarrow k_{i-1}$.
We re-build $TP$ (re-using the memory already allocated for it), and scan it left-to-right. For each maximal $TP[p,\dots,p+k-1]$ corresponding to a pair $ab$, if $ab$ is in $\mathcal H$ then, in order:
\begin{enumerate}
	\item we set $\mathcal H[ab]\leftarrow |B|$
	\item if $\mathcal F[k]\neq NULL$ we set the fifth component  of $B[\mathcal F[k]]$ ($Prev$ field) to $|B|$
	\item we append $\langle ab, p, k, k,NULL, \mathcal F[k] \rangle$ at the end of $B$
	\item we set $\mathcal F[k]\leftarrow |B|-1$ (decreased by one because we just increased $B$'s size).
\end{enumerate}

This procedure runs in $\bigO(n)$ time and uses $\bigO(1)$ words of space in addition to $T$, $TP$, $B$, $\mathcal H$, and $\mathcal F$.
Operations on $\qu$ are implemented as follows:

\begin{itemize}
	\item $\qu[ab],\ ab\in \extSigma^2$: return second, third, and fourth components of $B[\mathcal H[ab]]=\langle ab, P_{ab}, L_{ab}, F_{ab}, Prev_{ab}, Next_{ab} \rangle$. $\bigO(1)$ expected time.
	\item $\qu.max()$: return the first component of $B[\mathcal F[MAX]]$. $\bigO(1)$ time. 
	\item $\qu.min()$: return the first component of $B[\mathcal F[MIN]]$. $\bigO(1)$ time. 
	\item $\qu.remove(ab)$: delete $B[\mathcal H[ab]]$ from its linked list in $B$ and delete $ab$ from $\mathcal H$. If $ab$'s linked list in $B$ is now empty (i.e.\ $ab$ was the last pair with frequency $F_{ab}$), re-compute new MAX and MIN if necessary (by using $\mathcal F$'s list pointers), remove frequency $F_{ab}$ from the linked list in $\mathcal F$ and set $\mathcal F[F_{ab}]\leftarrow NULL$. $\bigO(1)$ expected time.
	\item $\qu.contains(ab)$. This operation requires one access to $\mathcal H$. $\bigO(1)$ expected time.
	\item $\qu.size()$: we only need to keep a variable storing the current size and update it at each remove operation. $\bigO(1)$ time.
	\item $\qu.decrease(AB)$: decrease the fourth component of $B[\mathcal H[AB]]$ (i.e.\ $F_{AB}$) by one. Remove $B[\mathcal H[AB]]$ from its linked list in $B$ and insert it in the list having $\mathcal F[F_{AB}]$ as first element (if $\mathcal F[F_{AB}]=NULL$, then create a new linked list and adjust $\mathcal F$'s list pointers). Re-compute MAX and MIN if necessary (by using $\mathcal F$'s list pointers). If $F_{AB}$ becomes equal to 1, remove the pair from the queue. $\bigO(1)$ expected time.
	\item $\qu.synchronize(AB)$. Let $\overline{ab}$ be a global variable storing the pair with the smallest frequency in $\qu$ (Algorithm \ref{alg:substitution round}, Line \ref{line:less freq pair}).
	We apply the sorting algorithm of Section \ref{sec: sorting pairs} to sub-array $TP'=TP[P_{AB},\dots,P_{AB}+L_{AB}-1]$, excluding text positions that contain a blank character. After this, positions are clustered in $TP'$ by character pairs. For each contiguous maximal sub-array  $TP[p,\dots, p+k-1]$ of $TP'$ corresponding to a pair $XY$:
	
	\begin{itemize}
		\item If $XY\neq AB$ and $k > F_{\overline{ab}}$, we remove $\overline{ab}$ and insert $XY$ in our priority queue as follows. Let $j_{min} = \mathcal{H}[\overline{ab}]$. We call $\qu.remove(\overline{ab})$, overwrite  $B[j_{min}] \leftarrow \langle XY, p, k, k, NULL, NULL \rangle$, insert $B[j_{min}]$ in the linked list having $\mathcal F[k]$ as first element (assigning a value to $Next_{XY}$) or create a new list if $\mathcal F[k]=NULL$, and insert $\mathcal H[XY] \leftarrow j_{min}$ in the hash. We re-compute MIN taking the least frequent pair between $XY$ and the pair corresponding to $B[\mathcal F[MIN]]$, and re-compute the minimum $\overline{ab}$ (i.e.\ the pair corresponding to $B[\mathcal F[MIN]]$).
		
		\item If $XY= AB$, then we just overwrite  $B[\mathcal H[AB]] \leftarrow \langle AB, p, k, k, Prev_{AB}, Next_{AB} \rangle$.
	\end{itemize}
	
	Running time: $\bigO(L_{AB})$, where $L_{AB}$ is $AB$'s interval length at the moment of entering in this procedure.
	
\end{itemize} 
\textbf{Time complexity} Since we insert at most $(\epsilon/13)\cdot n$ pairs in $\qu$ but there may be up to $\bigO(n)$ low-frequency pairs, once $\qu$ is empty we may need to fill it again with new low-frequency pairs (\texttt{while} loop at line \ref{line:while2} of Algorithm \ref{alg:repair}). We need to repeat this process  $\bigO(n/(n\cdot\epsilon/13))\in\bigO(1/\epsilon)$ times before all low-frequency pairs have been processed. Since operations at Lines \ref{line:sort pairs2}, \ref{line:new LF queue}, \ref{line:free mem2}, and \ref{line:compact text2} take $\bigO(n)$ time, the overall time spent inside these procedures is $\bigO(n/\epsilon)$. 
Using the same reasonings of the previous section, it is easy to show that the time spent inside $\qu.synchronize$ is bounded by $\bigO(n)$ thanks to our amortization policy. Moreover, since all queue operations except $\qu.synchronize$ take constant time, we spend overall $\bigO(n)$ time operating on the queue. 
These considerations imply that instructions in Lines \ref{line:while2}-\ref{line:compact text2} of Algorithm \ref{alg:repair} take overall $\bigO(n/\epsilon)$ randomized time. Theorem \ref{thm:main}(i) follows.\\\ \\
\textbf{Light queue} While for the fast queue we reserve $\epsilon\cdot n$ space for $B$ and $\mathcal H$, in the light queue we observe that we can re-use the space of \emph{blank text characters} generated after replacements. The idea is the following. Let $S_i$ be the capacity (in terms of number of pairs) of the queue at the $i$-th execution of the \texttt{while} loop at Line \ref{line:while2}; at the beginning, $S_1=1$.  After executing operations at Lines \ref{line:sort pairs2}-\ref{line:compact text2}, new blanks are generated and this space is available at the end of the memory allocated for the text, so we can accumulate it on top of $S_i$ obtaining space $S_{i+1}\geq S_i$. At the next execution of the \texttt{while} loop, we fill the queue until all the available space $S_{i+1}$ is filled. We proceed like this until all pairs have been processed. The question is: how many times the \texttt{while} loop at Line \ref{line:while2} is executed?


Replacing a pair $ab$  generates at least $F_{ab}/2$ blanks: in the worst case, the pair is of the form $aa$ and all pair occurrences overlap, e.g. in $aaaaaa$ (which generates 3 blanks). Moreover, replacing a pair with frequency $F_{ab}$ decreases the frequency of at most $2F_{ab}$ pairs in the active priority queue (these pairs can therefore disappear from the queue). Note that $F_{ab}\geq 2$ (otherwise we do not consider $ab$ for substitution).
After one pair $ab$ is replaced at round $i$, the number $M_i$ of elements in the active priority queue is at least $M_i \geq S_i - (1+ 2F_{ab})$. Letting $f_1,f_2,\dots$ be the frequencies of all pairs in the queue, we get that after replacing all elements the number ($0$) of elements in the priority queue is:
$$ 
0 \geq S_i -(1+2f_1) - (1+2f_2) - \cdots
$$
which yields
$$
S_i \leq (1+2f_1) + (1+2f_2) + \cdots
$$
since $f_i/2 \geq 1$ then $$ S_i \leq 2.5 f_1 + 2.5f_2 + \cdots =2.5 \sum_i f_i$$
so when the active priority queue is empty we have at least $\sum_i f_i/2 \geq S_i/5$ new blanks. Recall that a pair takes 13 words to be stored in our queue. In the next round we therefore have room for a total of $(1+1/(5\cdot 13))S_i=(1+1/65)S_i$ new pairs. This implies $S_{i} =  (1+ 1/65)^{i-1}$. Since $S_i\leq n$ for any $i$, the number $R$ of rounds can be computed as $S_R \leq n \Leftrightarrow (1+ 1/65)^{R-1} \leq n \Leftrightarrow R\in\bigO(\log n)$. 
With the same reasonings used before to analyze the overall time complexity of our algorithm, we get Theorem \ref{thm:main}(ii).

\section{References}
\bibliographystyle{IEEEbib}
\bibliography{repair}

\end{document}